# STEM analysis of deformation and B distribution in nanosecond laser ultra-doped Si$_{1-x}$ B$_x$

Géraldine Hallais*, Gilles Patriarche*, Léonard Desvignes*, Dominique Debarre*, Francesca Chiodi*,

* Université Paris-Saclay, CNRS, Centre de Nanosciences et de Nanotechnologies - C2N, Palaiseau 91120, France

## Abstract

We report on the structural properties of highly B-doped silicon (> 2 at. %) realised by nanosecond laser doping. We investigate the crystalline quality, deformation and B distribution profile of the doped layer by STEM analysis followed by HAADF contrast studies and GPA and compare the results to SIMS analyses and Hall measurements. When increasing the active B concentration above 4.3 at.%, the fully strained, perfectly crystalline, Si:B layer starts showing dislocations and stacking faults. These only disappear around 8 at.% when the Si:B layer is well accommodated to the substrate. When increasing B incorporation, we increasingly observe small precipitates, filaments with higher active B concentration and stacking faults. At the highest concentrations studied, large precipitates form, related to the decrease of active B concentration. The structural information, defect type and concentration and active B distribution are connected to the initial increase and subsequent gradual loss of superconductivity.

## I. Introduction: why study ultra-doped silicon?

When silicon is strongly doped with boron (>6x10$^{20}$ cm$^{-3}$), a superconducting phase appears in this covalent material [1,2,3]. To understand the mechanisms leading to the onset and evolution of superconductivity, the knowledge of the structural properties and of the boron distribution and activation is essential. Moreover, superconducting silicon devices are emerging [4,5], that need, to be designed and developed, such information on the homogeneity and activation of the doping. Due to the technical difficulties of realising such high doping levels, the insight on ultra-doped silicon material properties is still yet poor. In this paper, we discuss the results of an experimental study performed by STEM (Scanning Transmission Electron Microscopy), SIMS (Secondary Ion Mass Spectrometry) and transport Hall measurements on Si:B layers with doping in the 5.5x10$^{18}$ cm$^{-3}$ to 3.5x10$^{21}$ cm$^{-3}$ range realised by nanosecond laser annealing. To achieve such high doping, we employ an out of equilibrium technique, nanosecond laser doping, necessary to overcome the solubility limit (~$4 \times 10^{20}$ cm$^{-3}$) and to incorporate up to 12 at.% of B atoms [6,7]. When progressively increasing the doping, we observe an initial linear increase of the hole carrier density measured by Hall effect (i.e., the active B concentration, $n_{B\ Hall}$), followed by a slower non-linear increase and a decrease at higher doping (Fig.1b). At the same time, the superconducting critical temperature increases to a maximum corresponding to the maximum of the active B concentration, and then decreases. We thus expect the B active concentration to be a determining parameter for superconductivity. The aim of this paper is to understand the origin of this behaviour through the investigation of the structural properties and the boron distribution and activation in the three regimes described above (linear, non-linear, decreasing) by means of complementary STEM, SIMS and electrical analyses.

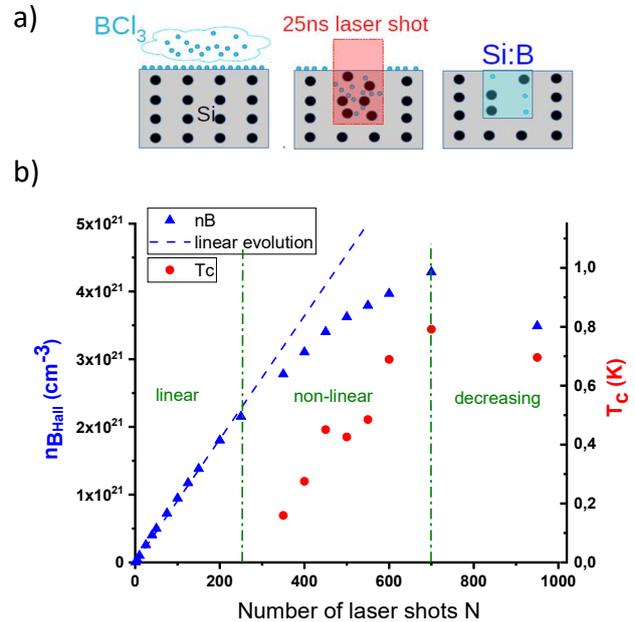

*Figure 1. (a) GILD process: chemisorption of the precursor gas over the sample surface; melting of the substrate by the laser pulse and B introduction in the liquid Si; fast cooling and epitaxy of a Si:B layer on top of the Si substrate. (b) (Left) active B concentration measured by Hall effect and (right) superconducting critical temperature as a function of the number of laser shots for a 176 nm thick Si:B layer.*

In addition, we extract the dopant distribution within the layer from STEM images either from the lattice deformation or the High Angle Annular Dark Field (HAADF) intensity, and comment on the precision and limitations of these analysis.

## II. Ultra-doping: Gas Immersion Laser Doping

Gas Immersion Laser Doping (GILD) [8] (Fig.1a) is performed in an Ultra High Vacuum reactor (~ $10^{-9}$ mbar) which insures a very low impurity level. A puff of the boron precursor gas, pure $BCl_3$, is injected using a pulse valve onto the (100) oriented high resistivity n-type silicon surface to induce a pressure of ~$10^{-5}$ mbar, just enough to saturate the chemisorption sites. As the gas is continuously pumped, photolytic or pyrolytic CVD processes are avoided, and the supplied quantity of dopant atoms, from the chemisorbed layer, is constant and self-limited. After a small delay, a pulse of an excimer XeCl laser ($\lambda = 0.308$ µm with pulse duration 25 ns) is sent to the sample area and absorbed over ~ 7 nm. Given the electron-phonon characteristic time, the light energy can be considered as completely and instantly converted into thermal energy. The heat produced is evacuated very quickly (87mm$^2$.s$^{-1}$) [9] by one-dimensional diffusion to the substrate (in 25 ns the heat diffuses over about 1µm). The thermal energy melts the silicon from its surface, the silicon melting threshold being 600 mJ.cm$^{-2}$. The greater the amount of laser energy absorbed, the deeper the melting front advances in the substrate. Thus, the density of laser energy is proportional to the doped thickness. All chemical species diffuse in the liquid. Once the thermal energy is dissipated in the substrate and the local temperature decreases below the melting temperature, an epitaxial recrystallization front rises to the surface at a speed of about 4 m/s [10], slow enough for the crystal to reconstruct from the underlying crystal lattice in the absence of defects (epitaxy) and fast enough to trap in substitutional sites the boron atoms with a segregation coefficient close to 1, achieving concentrations larger than the solubility limit. When the crystallization front reaches the surface, the excess impurities contained in the liquid are expelled outwards, such as Cl whose segregation coefficient is close to 0 [9]. To improve the uniformity of the 2mm x 2mm laser spots, and thus achieve a straight interface between the doped layer and the substrate, the spatial inhomogeneity of the laser energy density is reduced to about 1% by a careful optical treatment of the laser beam using, in particular, a fly-eye homogenizer (a system composed of 2 squares 11x11 micro lens matrices).

This entire chemisorption-melting-crystallisation process can be repeated the desired number of times (number of laser shots N) to increase the overall concentration of dopants in a single layer, as shown in Fig.1b.

## III. Measuring the Si:B structure and the B distribution

In this paper, we prepared several samples by fixing the doped thickness to 176nm (value measured by SIMS and STEM) and varying the number of shots from 1 to 950, tuning the hole concentration from $5.5 \times 10^{18}$ to $3.8 \times 10^{21}$. We present the sample analysis results for 200 laser shots (linear regime), 350 and 700 (non-linear regime), and 950 shots (decreasing regime).

*STEM Analysis*

The technical details concerning the STEM measurements are specified in the Methods section. STEM observations enable us a precise analysis of the crystalline quality of the Si:B layer epitaxial on the Si substrate. Crystalline defects such as dislocations or stacking faults can thus be observed. In this work, we will present only the images obtained in HAADF mode (Fig. 2).
In addition, on the same images, we performGeometrical Phase Analysis (GPA), an analysis in the Fourier space, ([11,12]) which provides a powerful tool to measure the deformations quantitatively (Fig.3). The Si:B layer in-plane deformation ($\varepsilon_\parallel$) and out-of-plane deformation ($\varepsilon_\perp$) are thus measured in relation to the Si substrate:

$$\varepsilon_\parallel = \frac{a_{SiB\parallel} - a_{Si}}{a_{Si}} \qquad \varepsilon_\perp = \frac{a_{SiB\perp} - a_{Si}}{a_{Si}} \qquad (1)$$

where $a_{Si} = 0.5431$ nm.

*Extraction of the B distribution profile*

The concentration profiles cannot be directly extracted from STEM-EDX analysis since the energy of the K$\alpha$ ray of the B is too close to the L$\alpha$ line of the Si (spectrum Fig.4a). In addition, the peak intensity BK$\alpha$ is very low. We will nevertheless benefit from EDX to estimate qualitatively the evolution of the layer concentration at the highest doping. We can instead extract the dopant profiles by two techniques, either through the lattice deformation only (GPA), or through both the lattice deformation and the HAADF contrast.
In the first case, we exploit Vegard's law, stating that in a binary compound, the lattice parameter is directly related to the composition by:

$$a_{SiB} = n_{B\%} a_B + (1 - n_{B\%}) a_{Si} \qquad (2)$$

To extract the Si:B lattice parameter in the hypothesis of an isotropic material, one can use Si Poisson coefficient to relate the perpendicular and parallel deformations measured by GPA. The resulting concentration is given by:

$$n_{B\ vegard}(\%) = \frac{a_{Si}}{(1-K)(a_B - a_{Si})} (K\varepsilon_\parallel - \varepsilon_\perp) \qquad (3)$$

Where $K = -\frac{2\nu}{(1-\nu)} = -0.77$ for the silicon value $\nu = 0.278$. Further details on eq.3 are given in the methods section. As eq.3 shows, to extract $n_{B\ Vegard}$ it is necessary to assume a value for the lattice parameter of B, $a_B$, in a cubic lattice, which is not the case. In the literature, values range from $a_B = 0.378$ nm [14] to $a_B = 0.4084$ nm [15]. Thus, $a_B$ and the Si:B Poisson coefficient $\nu$ are the main sources of uncertainty in the derivation of the substitutional B concentration profile from the lattice in-plane and out-of-plane deformations.

HAADF images can also be exploited to deduce the B concentration profile. Indeed, the high angular integrated intensity is directly related to the atomic number Z of the atomic columns scanned by the electron probe [13]. If we assume that in the binary compound Si:B, all the B atoms replace Si atoms in the crystal lattice, and that there aren't other elements present, it is theoretically possible to calculate the B concentration from the deformation and the HAADF contrast, defined as the ratio between the layer intensity ($I_{SiB}$) relative to the volume of Si:B crystal lattice ($V_{SiB}$) and the substrate intensity ($I_{SiB}$) relative to the volume of the Si crystal lattice ($V_{Si}$):

$$C_{HAADF} = \frac{I_{SiB}/V_{SiB}}{I_{Si}/V_{Si}} \tag{4}$$

The B percent atomic concentration, $n_{B\%}$ is given by eq.5 (see also Methods section for further details)

$$n_{B\ HAADF}(\%) = \frac{[(1-\varepsilon_\perp)(1-\varepsilon_\parallel)^2 C_{HAADF} - 1]}{Z_B^{1.7} - Z_{Si}^{1.7}} \tag{5}$$

We stress that this expression applies to monocrystalline layers without defects or impurities.

*SIMS Analysis*

SIMS analysis was performed by taking special care to the quantification of B doping for the studied concentrations, higher than $10^{21}$ cm$^{-3}$. In particular, an oxygen primary beam was employed to correct the matrix effects, and only secondary ions at high energy (> 100 eV) were analysed as they are less sensitive to the chemical surrounding (see Methods section for further details). This technique makes it possible to quantify the total amount of B ($n_{B\ SIMS}$) present in the doped layer regardless of his position in the crystal.

*Hall transport measurements*

The hole concentration was measured by Hall transport on the same spots analysed by STEM. Hall measurements were performed on a Hall cross etched on each spot with a central region 300µm x 300µm. The influence of the substrate is negligible due to the n-p barrier between the p-type layer and the n-type substrate. The hole concentration is directly related to the active B concentration as each B atom provides a hole carrier:

$$n_{B\ Hall} = \gamma \frac{1}{dqR_H} \tag{6}$$

Where q is the electron charge, d is the thickness of doped layer, $R_H$ is the Hall coefficient and $\gamma = 0.75$ [16] is the Hall mobility factor, the ratio between the Hall mobility $\mu_H$ and the conductivity mobility $\mu_c$: $\gamma = \mu_H/\mu_c$. The main uncertainty on $n_{B\ Hall}$ is related to the Hall mobility factor, which varies in the literature between $\gamma=0.7$ and $\gamma=0.8$, and whose value has not been measured above $10^{21}$ cm$^{-3}$.

## IV. Results and discussions

In the following part, we describe the Si:B layer properties in the three regimes of active B evolution with the number of laser shots N (linear increase, non-linear increase and decrease, see Fig.1). For each regime we will summarise the results obtained with all the characterisation methods employed: STEM-HAADF images, to observe the crystalline quality of the layer, its structural defects, and B precipitates (Fig.2); GPA analysis, to observe the in-plane and out-of-plane lattice deformation (Fig.3); STEM-EDX images to compare the amount of B in the layer and the precipitates (Fig. 4); and the B concentration profile extracted from SIMS, Hall, and/or STEM analysis through Vegard's law or HAADF contrast, to observe the dopant homogeneity and accumulation (Fig.5).

*Linear regime*

A characteristic example of the behaviour in the linear regime is the 200 laser shots sample. The Si:B layer presents the same crystal lattice quality as the Si substrate (Fig.2a) with no defects. The inset Fast Fourier Transform (FFT) pattern reveals an epitaxial orientation relationship (cube-on-cube) between the two lattices.

Fig.3a shows that the crystal lattice of the Si:B layer is tensile strained, with an out-of-plane deformation $\epsilon_\perp$ roughly constant within the layer, varying from $\epsilon_\perp = -1.5\ \%$ in the 35 nm above the Si:B/Si interface, to $\epsilon_\perp = -1\%$ at the surface (Table 1). No in-plane deformation $\epsilon_\parallel$ is observed. This is as expected when the doping is low enough that a monocrystalline Si:B layer can be epitaxied without defects on the Si substrate through elastic relaxation. Indeed, due to the epitaxial relation with the substrate, the in-plane lattice constant is equal to the Si one while the out-of-plane lattice constant is reduced to accommodate for the smaller lattice volume due to the smaller B size. From the deformation profile, we extract the concentration profile of the substitutional B, $n_{B\ vegard}$, with eq.3. The profiles thus obtained with $a_B$ in the range given by literature (Fig.5a, blue lines) show $n_{B\ vegard}=1.6\times10^{21}$ cm$^{-3}$ constant over approximately 40nm above the Si:B/Si interface, followed by a roughly linear decrease down to $1.05\times10^{21}$ cm$^{-3}$ at the surface. The SIMS profile $n_{B\ SIMS}$, sensitive to the total amount of B, is similar but with higher values of concentration, $n_{B\ SIMS} = 2.3\times10^{21}$ cm$^{-3}$ near the interface to $1.9\times10^{21}$ cm$^{-3}$ near the surface (Fig.5a green line). The difference between the two is possibly due to the presence of non-electrically active B ($n_{B\ inactive}$), $n_{B\ inactive}=n_{B\ SIMS}-n_{B\ vegard}$. The value obtained by Hall measurement, $n_{B\ Hall}$, sensitive only to active, substitutional B atoms, confirms this hypothesis: $n_{B\ Hall} \sim n_{B\ vegard} < n_{B\ SIMS}$, the agreement being particularly good in the bottom 40 nm.

Even though SIMS, Hall measurements and the deformation all draw a coherent picture, HAADF is puzzling. Indeed, we observe a HAADF contrast $C_{HAADF} >1$: the intensity of the Si:B layer is larger than that of the substrate (Fig.3a), while we would expect $C_{HAADF} <1$ for substitutional B atoms (eq.4) since $Z_B$ (=5) < $Z_{Si}$ (=28). It is thus impossible to calculate the concentration profile B by eq.5. This gives us an indication that another phenomenon is masking the replacement of Si atoms by B atoms in the crystal lattice, such as the presence of substitutional atoms with a larger Z than the Si. However, the EDX spectra on all Si:B layers (see e.g. Fig.4a for samples with 700 and 950 laser shots) do not show any other elements. The most likely hypothesis is the homogeneous and significant presence of Si interstitial atomsI It is unclear how to explain such important presence of interstitial Si. We note that the same result was obtained for a second lamella on the same spot and in other spots at doping N=350.

*Non-linear regime*

When increasing the doping to attain the non-linear regime, (characterised by the N=350 laser shots sample) we observe the appearance of dislocations at the Si:B/Si interface and the propagation of stacking faults in the thickness of the layer that show the beginning of plastic relaxation (Fig.2b).

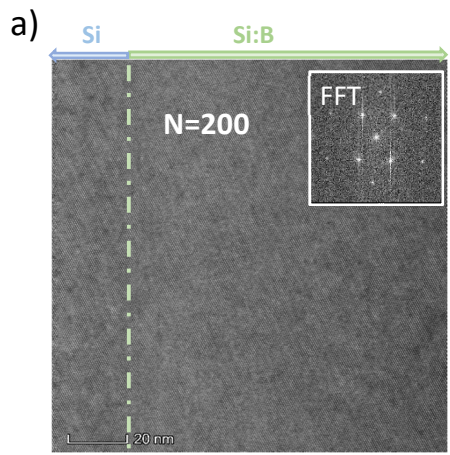

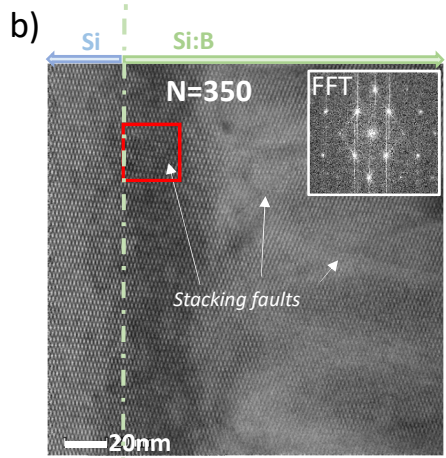
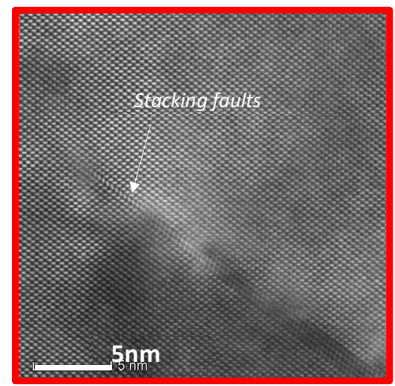

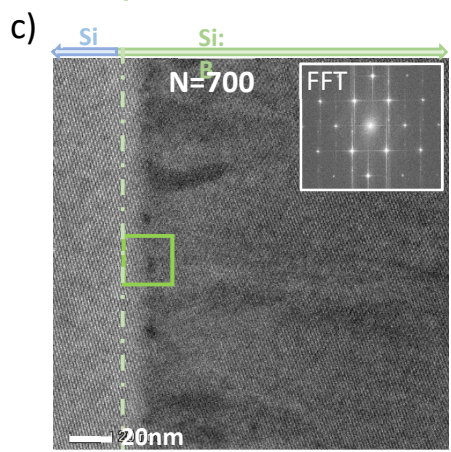
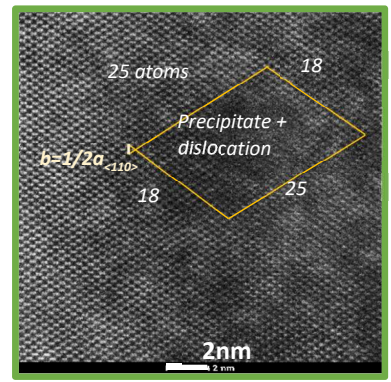

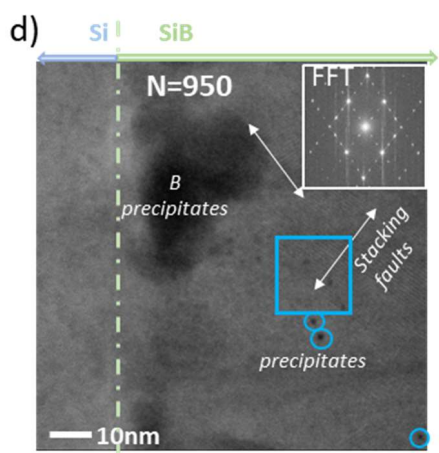
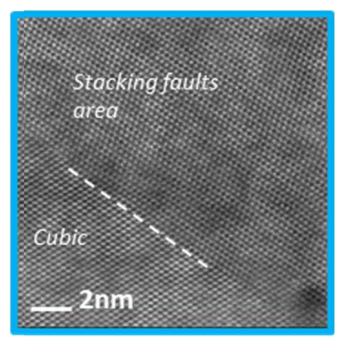
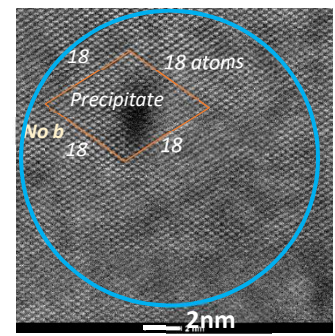

*Figure 2. STEM-HAADF images of the doped layer with a) N=200, b) N=350, c) N=700 and d) N=950 shots.*
*The FFT of each image on the Si:B part gives indications on the crystal quality. On the N=350, N=700 and N=950 shots images, the inserts are focused on the crystal defects (stacking faults and dislocations with burgers vector determination) or on the precipitates.*

These stacking faults and dislocations also appear in the inset FFT pattern as tilted lines connecting the diffraction spots, as well as in the GPA charts (Fig.3b). At the Si:B/Si interface, we observe a fully strained 15 nm layer with $\epsilon_\perp$ =-1.3% and $\epsilon_\parallel$ =0(Fig.3b). Above it, a second fully strained layer of about 35 nm has a maximum $\epsilon_\perp$ =-2.6%. The dislocations start from the bottom of these two layers. Finally, the rest of the Si:B layer (130nm) has a constant $\epsilon_\parallel$ =-0.7% and $\epsilon_\perp$ decreasing to -2%. Indeed, the doping is too high to build a fully strained, monocrystalline layer, and three layers form within the doped region: the first just above the Si substrate, fully strained, correspond to the low doping concentration of the interface, the second is as thick as a monocrystalline layer can be epitaxied at the concentration of the liquid phase (~$3 \times 10^{21}$ cm$^{-3}$), while the last layer to crystallize is partially relaxed, with an important increasing in-plane deformation and a decreasing out-of-plane deformation when proceeding upwards.

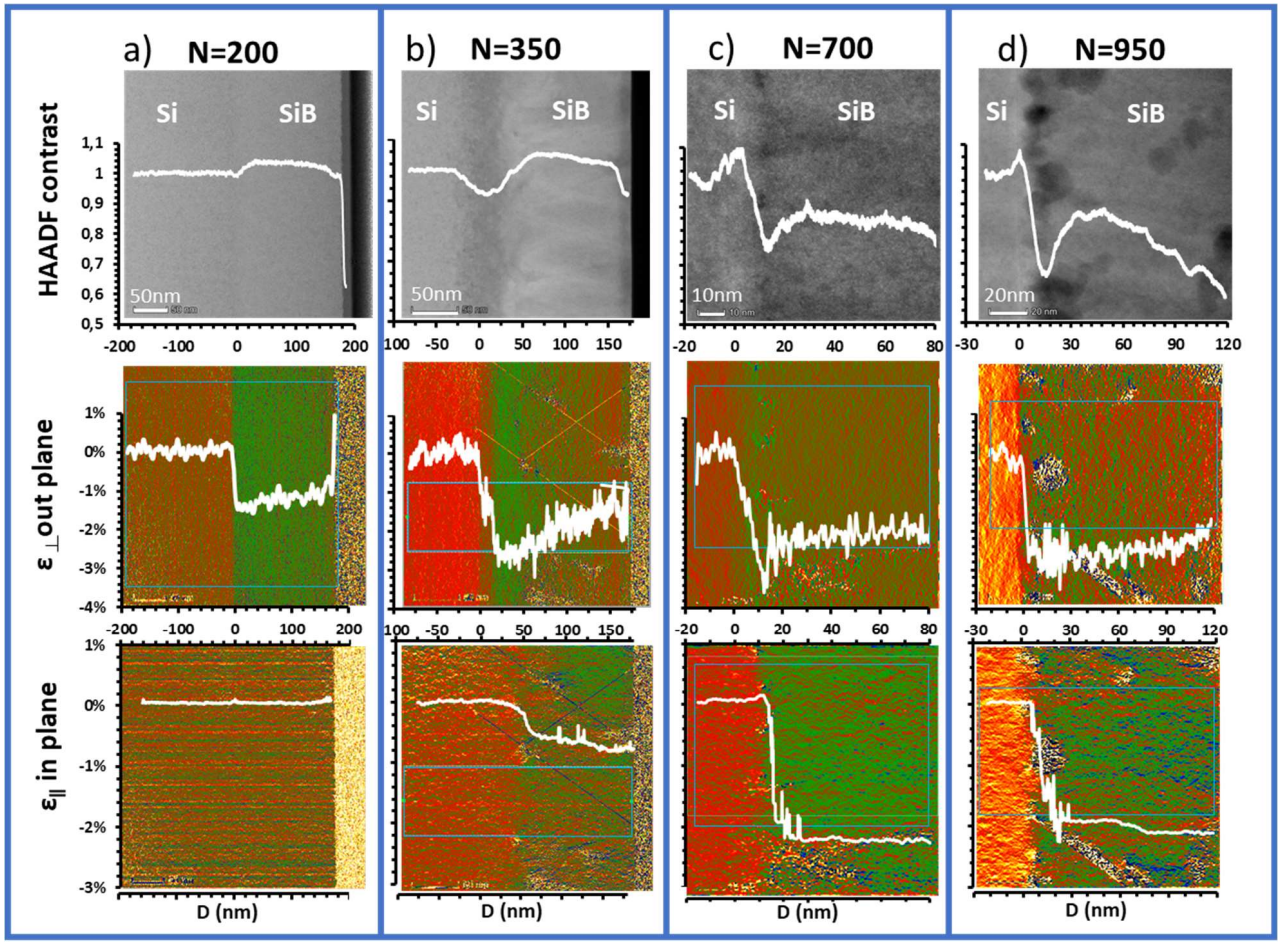

*Figure 3. STEM-HAADF images of the doped layer with a) N=200, b) N=350, c) N=700 and d) N=950 shots. Below each HAADF image, are represented the out-of-plane and in-plane deformation maps with the averaged profiles.*

Fig.5b (blue lines) shows the substitutional B concentration profile extracted with Vegard's law. We observe $n_{B\ vegard}$ =2.6x10$^{21}$ cm$^{-3}$ over approximately 70 nm above the Si:B/Si interface, in good agreement with the value given by Hall measurements. The SIMS profile (Fig.5a, green line) gives as before a total concentration larger than the Hall/Vegard's law one with a similar profile. Reasoning as for N=200, we could interpret the difference between $n_{B\ SIMS}$ and $n_{B\ Vegard}$ by an amount of inactive B. However, at N=350 $n_{B\ inactive}= n_{B\ SIMS}$- $n_{B\ vegard}$ is smaller than at N=200, pointing towards a smaller concentration of interstitial B. As we expect the incorporation of inactive B to be monotonically increasing with doping, we should instead obtain $n_{B\ inactive}$ (N=200) < $n_{B\ inactive}$ (N=350). A possible interpretation is that, at N=200, the deformation is sensitive to the amount of interstitial Si (assumed to explain the $C_{HAADF}$>1), and thus $n_{B\ inactive}$, preventing a quantitative extraction of $n_{B\ Vegard}$ for N=200. This may also be true for N=350, to a lesser extent. Indeed, for N=350 as for N=200, the HAADF contrast is >1 in the upper 150 nm of the layer. In the bottom layer we find as expected a HAADF contrast <1, which however does not allow to extract a reasonable B concentration (4.6 x$10^{21}$ cm$^{-3}$), possibly due to a distortion caused by small precipitates.

At N=700 shots (Fig.2c), we reach the end of the non-linear regime. HAADF images show dislocations concentrated at the Si:B/Si interface, and we recover a good crystalline quality in the doped layer, similar to that of the untreated substrate, with no stacking faults. The Si:B layer has thus well accommodated its lattice on the substrate by plastic deformation. The insert FFT pattern is similar to the one obtained on the sample images with N=200. We observe, as before, the formation of three layers (Fig.3c). The first, at the interface, is fully strained, with a small out-of-plane deformation over 5 nm and corresponds to the doping decrease at the end of the Si:B layer. The second, 12 nm thick, is also fully strained with no in-plane deformation and $\epsilon_\perp$ increasing up to the maximum $\epsilon_\perp$ =-3.5%. The rest of the layer is nearly fully relaxed with $\epsilon_\parallel$ ~-2.1% and a similar $\epsilon_\perp$ =-2%. At the interface between the first and second, fully strained layers, we observe small dark areas corresponding to small precipitates and misfit dislocations identified by their Burger vector 1/2$a_{Si}$ [110] (Lomer loops). Note that the thickness of the fully strained layer decreases gradually when B active concentration is increased, from the whole layer, 176 nm, at $n_B$=1.6x$10^{21}$ cm$^{-3}$, to 35 nm at $n_B$~3x$10^{21}$ cm$^{-3}$, and finally to 12 nm at $n_B$~4x$10^{21}$ cm$^{-3}$. This is in agreement with the expected pseudomorphic sublayer thickness which decreases as ~1/$n_B$.

On the HAADF image (Fig.2c), we can also notice column-shaped regions spanning up to the thickness of the layer, of dark contrast and therefore less rich in Si. Presumably they are richer in B. This may be explained as a beginning of the cellular breakdown process [17], where the liquid-solid interface roughness leads to lateral impurities segregation with respect to the main solidification front, and forms filamentary crystalline features where a large dopant concentration can be found in substitutional sites [18].

On Fig.5c (blue lines), the B concentration profile obtained by the Vegard's law shows an almost constant value on the first 80 nanometres of the Si:B layer from the interface. These values are in good agreement (between 89 and 92% depending on $a_B$) with the Hall effect measurement. The concentration measured by SIMS (Fig.5c, green line) is about 34% higher than $n_{B\ Vegard}$ or $n_{B\ Hall}$. The difference between the SIMS profile and Vegard's profile is explained as inactive B atoms that do not participate to the deformation of the Si:B lattice. The larger quantity of inactive B as compared to N=350 follows the expected evolution. We observe an increase of the inactive B concentration at the Si:B/Si interface, present also at N=350, which may be associated to the small precipitates observed by HAADF (Fig.2c).

The profile $n_{B\ HAADF}$ obtained from $C_{HAADF}$ by eq.5 corresponds to the profile given by the Vegard's law and the Hall measurement value in the layer ~40 nm above the interface. On the other hand, at the interface the concentration extracted from the HAADF is larger than the expected $n_{B\ HAADF}$ ~ $n_{B\ Vegard}$ due to the effect of the small precipitates. This is confirmed by the similar profiles of SIMS and HAADF at the bottom of the layer.

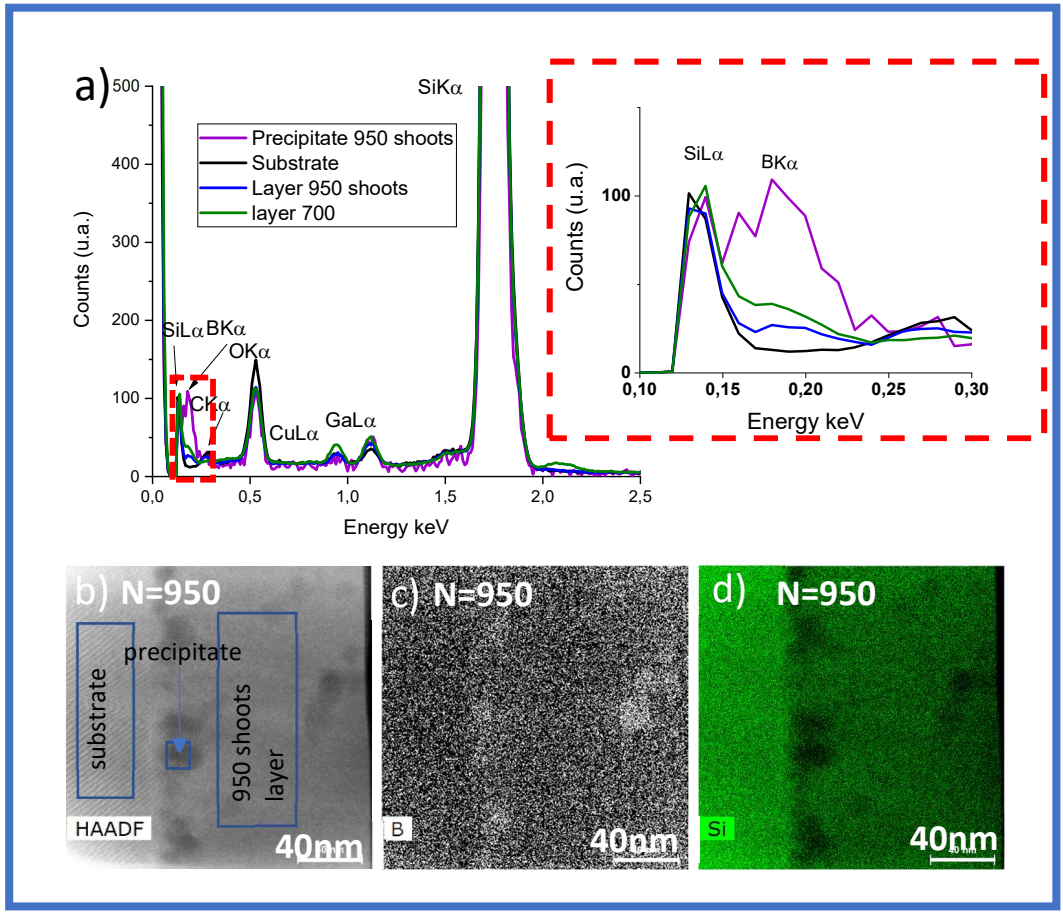

*Figure 4. STEM-EDX a) spectrums, c) and d) respectively B and Si mappings corresponding at the HAADF image b) on a Si:B layer with 950 shots. On the spectrums, the C and O peaks are due to surface contamination of the FIB section, the Ga peak is due to the preparation of the thin lamella with a $Ga^+$ ions beam, the presence of the Cu peak is due to the fluorescence on the TEM grid.*

*Decreasing regime*

Finally, the highest B concentration studied (N=950 shots) shows a decrease of the active concentration. Large black areas are observed in HAADF contrast at the Si:B/Si interface. An EDX map (Fig.4 b,c,d) of the layer shows that in these dark areas the lack of Si corresponds to B accumulation (spectrum Fig.4a, purple line). These B precipitates are amorphous since the crystallinity is lost in the FFT. Large (>30nm) micro maculated areas extend from these precipitates towards the surface (Fig.2d). Their signature is also visible on the FFT pattern as extra points (FFT Fig.2d). Furthermore, we observe small precipitates through the layer (black dots in Fig.2d). They are incoherent with the Si matrix, since the Burger vector is zero, indicating the absence of global dislocations (i.e zoom Fig.2d).

Similar to the previous samples, the layer presents three sub-layers (Fig.3d). The first, ~5 nm thick, is fully strained with $\epsilon_\perp$ increasing to $\epsilon_\perp$ =-2.5 and no in-plane deformation. The second, ~20nm thick, is partially relaxed with $\epsilon_\perp$ =-2.8% and $\epsilon_\parallel$ increasing up to $\epsilon_\parallel$ =-2%, while the third, fully relaxed layer shows $\epsilon_\perp$ slowly decreasing up to $\epsilon_\perp$ =-2% and $\epsilon_\parallel$ increasing in two steps to $\epsilon_\parallel$ =-1.9% and -2.1%.

On Fig.4d blue line, the B concentration calculated with the Vegard's law is almost uniform and is in good agreement (between 97% and 84% depending on $a_B$) with the Hall measurement. The value calculated with the $C_{HAADF}$ is much higher than the Vegard's value (41%) which suggests that the HAADF contrast is distorted by the presence of the many small precipitates in the layer and the large precipitates mostly found at the interface (see the peak in the HAADF) and in the thickness of the layer.

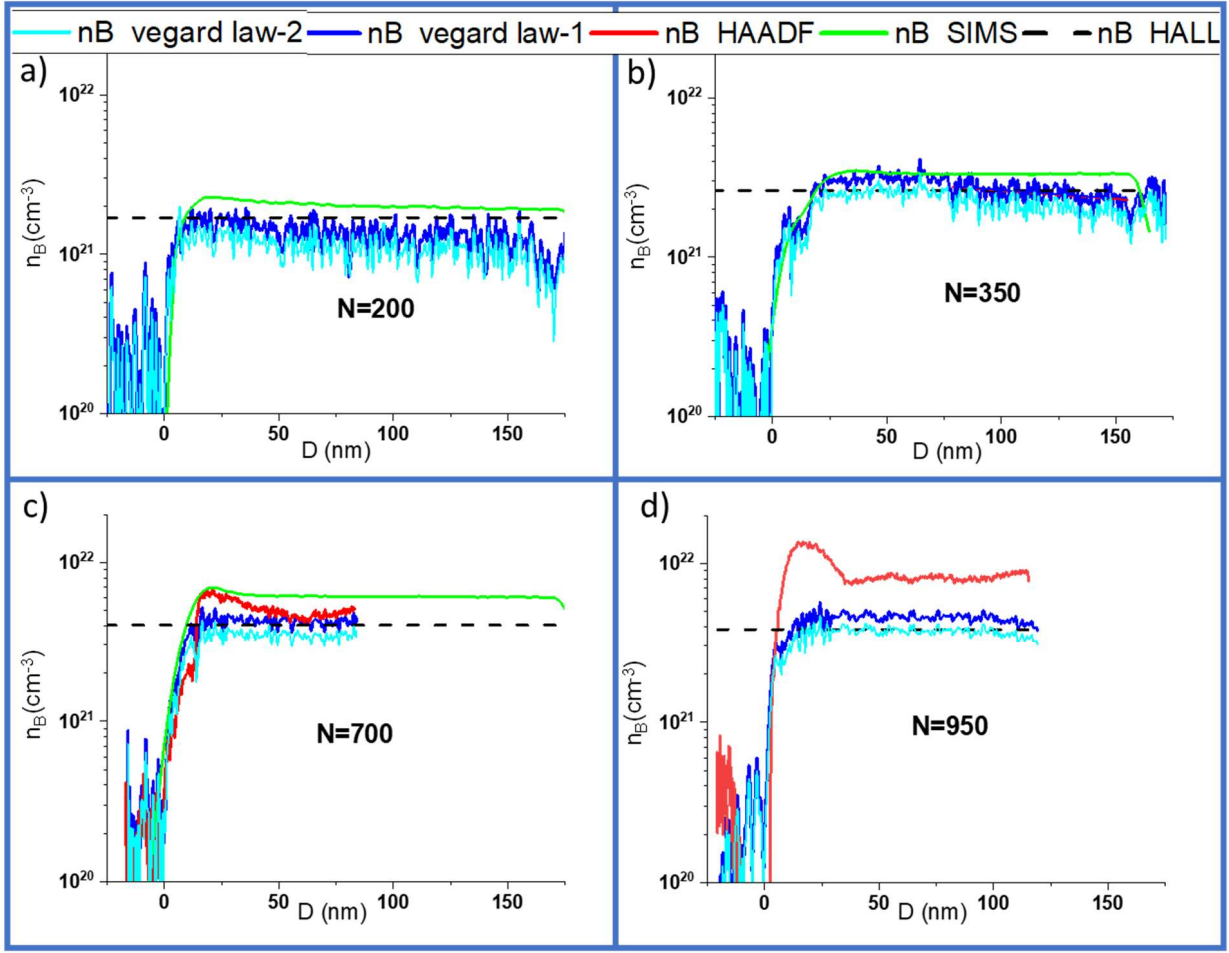

*Figure 5. B concentration profiles a) N=200, b) N=350, c) N=700, d) N=950 measured by SIMS (green line), calculated from Hall measurements (black dashed line), from deformation profiles (blue line dark "Vegard's law-1" with high $a_B$, light-blue line "Vegard's law-2" with a low $a_B$) and from HAADF contrast (red line)*

| | | N=200 | | N=350 | | N=450 | | N=700 | | N=950 | |
|---|---|---|---|---|---|---|---|---|---|---|---|
| | | SiB/Si interface | SiB layer | SiB/Si interface | SiB layer | SiB/Si interface | SiB layer | SiB/Si interface | SiB layer | SiB/Si interface | SiB layer |
| GPA deformation | thickness (nm) | 35 | 141 | 15 | 30 | 131 | 35 | 141 | 17 | 75 studied | 5 | 15 | 120 studied |
| | $\varepsilon_\perp$ (%) | -1,42 | -1,3 ↘ -0,7 | -1,31 | -2,59 | -2,5↘-1,2 | -2,5 | -2,2 | ↗ -3,4 | -2 | ↗ -3 | -2,8 | ↘2 |
| | $\varepsilon_\parallel$ (%) | 0 | 0 | 0 | -0,1 ↗ -0,6 | -0,6 ↗ -0,7 | +0,4 ↗ -0,91 | -2,18 | 0 | -2,18 | 0 | -1,9 | ↗ 2,1 |
| $n_B$ (cm$^{-3}$) × 10$^{-21}$ | thickness (nm) | 40 | 136 | 70,00 | 106 | 20 | 156 | 25 | 67 studied | 35 | 120 studied |
| | HALL | 1,7 | | 2,60 | | 3,4 | | 4,04 | | 3,86 | |
| | Vegard law | 1,6 / 1,3 | 1,4 ↘ 1,1 / 1,1 ↘ 0,9 | 3,15 / 2,57 | 2,8 ↘ 2,3 / 2,1 ↘ 1,7 | 0 ↗ 2,8 / 0 ↗ 2,3 | 2,8 / 2,3 | 4,4 / 3,59 | 4,25 / 3,46 | 3,8-4,6 | |
| | $C_{HAADF}$ | — | — | 4,60 | — | — | — | 6,7 | 5 | 13,3 max | 8,1 |
| | SIMS | 2,3 | 2,1 ↘ 1,9 | 3,50 | 3,3 | 4,4 | 4,3 | 6,7 | 6,1 | — | |

*Table 1. Summary of some deformation's values measured by GPA analysis and B concentration measured or calculated by different methods (SIMS, Hall measurements, Vegard's law, $C_{HAADF}$).*

## V. Conclusions.

Through complementary analyses (STEM HAADF-GPA, SIMS, Hall), we investigate the phenomena that occur when we incorporate high B concentrations as high as $n_B$=12 at.% into silicon <100> by nanosecond laser doping. At B concentrations lower than $n_B$=4.3 at.%, active B varies linearly with the number of laser shots. This is expected from the constant incorporation of dopants at each laser shot, determined by the self-limited precursor gas concentration chemisorbed at the sample's surface. In this case, the majority of B is found in substitutional positions in the Si crystal lattice. The Si:B layer presents an elastic relaxation in the growth plane that doesn't induce crystalline defects, forming a fully strained layer with deformation up to $\epsilon_\perp$ =-3%. SIMS analysis gives a higher concentration than that obtained by the Vegard's law or by Hall measurement: at $n_B$=3 at.%, 15% of the incorporated B is electrically inactive and doesn't participate in the deformation of the crystalline lattice. Moreover, an amount of interstitial Si needs to be assumed to understand the STEM-HAADF analyses, while no other contamination was detected by STEM-EDX at the detection limit (0.1at. %). Between 4.3 at.% and 8 at.%, the increase in active B concentration begins to gradually slow down with the number of laser shots, while the inactive B concentration grows to 34%. We observe the beginning of a plastic relaxation with the formation a fully strained sublayer a few tens of nm thick at the Si:B/Si interface and a partially relaxed sublayer on top, with increasing deformation in and perpendicular to the layer plane. This incomplete relaxation generates many crystalline defects, such as dislocations and stacking faults in the layer. We also still observe a small amount of interstitial Si above the thin fully strained layer. Surprisingly, superconductivity appears in this relaxed phase. At the end of this non-linear regime, around 8 at.% active B, the plastic relaxation is complete in the sublayer above the first fully strained sublayer. The lattice crystal is well accommodated to the substrate with only localized misfit dislocations at the Si:B/Si interface, where few B precipitates begin to appear. We also observe inhomogeneities in the active B concentration as filaments starting from the interface with the fully strained layer possibly related to cellular breakdown processes. According to the HAADF contrast there is no more Si in interstitial positions. When the number of shots is increased further, the concentration of active B decreases. Above the 5 nm thin fully strained layer, the lattice is completely relaxed, and B creates large precipitates mostly at the Si:B/Si interface, and a few in the layer. The STEM observation of a N=950 shots sample shows that the lattice crystal is completely at fault with large twinned domains. A crystalline disorder sets in, which decreases the electrical and in turn the superconductive performances. It should be also noted that despite the uncertainty linked to the parameter $a_B$ which varies in the literature by ~ 8%, the calculation of the active B concentration profile with the deformations measured by STEM-GPA and the Vegard's law is consistent with the values measured by Hall effect. The calculation seems to give better results if we adopt a smaller $a_B$ for disturbed regimes with many dislocations and stacking faults (N=350: start of plastic relaxation and N=950) and a larger $a_B$ for stable regimes without crystalline defects in the Si:B layer (N=200: elastic relaxation and N=700: plastic relaxation). Obtaining the B concentration profile by studying the HAADF contrast is possible (i.e. N=700) but the presence of interstitial impurities and B precipitates disturbs the contrast.

## VI. Methods / Technical details

*STEM Analysis*

For the STEM analysis, a thin lamella (<100nm) is machined vertically in the laser-doped spot thanks to a Focus Ions Beam using a FEI ThermoFisher SCIOS dual beam SEM (UHR NiCol) / FIB (Siderwinder 550V-30kV) with an in-situ Easy-lift micromanipulator. All samples were observed in an aberration-corrected FEI ThermoFisher TEM/STEM TitanThemis 200 operating at 200 keV. The convergence half-angle of the probe was 17.6 mrad and the detection inner and outer half-angles for HAADF-STEM were 69 mrad and 200 mrad, respectively. The lamella was imaged along the ⟨1 1 0⟩ zone axis. All micrographs where 2048 by 2048 pixels. The dwell time was 8 µs and the total acquisition time 41s. GPA is performed in Digital Micrograph software on STEM-HAADF images. Energy Dispersive X-ray (EDX) measurements were performed in the Titan microscope featuring the Chemistem system, that uses a Bruker windowless Super-X four-quadrant detector and has a collection angle of 0.8 sr. The acquisition time for the mappings was 10min, during which no significant drift occurred. This analysis allows us to know the elements present in the sample at a concentration greater than or equal to 0.1 at.%.

*SIMS Analysis*

SIMS measurements were performed with a 4F Cameca system equipped with a magnetic mass spectrometer. For doping levels above $5 \times 10^{20}$ cm$^{-3}$, the doping value can be difficult to extract due to matrix effects. However, it was shown that using an oxygen primary beam, the variation of the relative ion yield is the same for B and Si, so that the matrix effect can be corrected by the comparison with the silicon signal and using the Relative Sensibility Factor (RSF) [12]. Moreover, only the secondary ions at high energy (> 100 eV) were used for the analysis, as they are less sensitive to the chemical surrounding. Boron concentration is then obtained by comparison with a standard from the National Institute of Standards (NIST). Depth calibration is obtained with a mechanical profilometer and we consider the sputter rate difference between the doped layer and the silicon substrate.

*Extraction of the dopant profile*

- *Vegard's law*

It is possible to calculate the profile of the B concentration from the GPA measurements. Indeed, in a binary compound, the lattice parameter is directly related to the composition by the Vegard's law (eq.2).

$$a_{SiB} = n_{B\%} \, a_B + (1 - n_{B\%})a_{Si} \qquad (2)$$

$$n_B \, (\%) = \frac{a_{SiB} - a_{Si}}{a_B - a_{Si}}$$

In an isotropic material, the perpendicular deformation is related to the parallel deformation by the Poisson coefficient:

$$\frac{a_{SiB\perp} - a_{SiB}}{a_{SiB}} = K \frac{a_{SiB\parallel} - a_{SiB}}{a_{SiB}}$$

With $K = -\frac{2\nu}{(1-\nu)} = -0.77$ if we adopt the Poisson coefficient of Si ($\nu = 0.278$) which however may evolve in Si:B at higher concentrations.

$$a_{SiB} = K \frac{a_{SiB\perp} - K a_{SiB\parallel}}{1 - K}$$

Thus, starting from the deformations measured through the GPA (eq.1), it is possible to extract $a_{Si:B}$ then $n_B$ though eq.3:

$$n_{B \, Vegard} \, (\%) = \frac{a_{Si}}{(1-K)(a_B - a_{Si})} (K\varepsilon_\parallel - \varepsilon_\perp) \qquad (3)$$

- *HAADF contrast*

The HAADF contrast is defined as the ratio between the intensity of the layer ($I_{SiB}$) relative to the volume of Si:B crystal lattice ($V_{SiB}$) and the intensity of the substrate ($I_{SiB}$) relative to the volume of Si crystal lattice ($V_{Si}$) (eq.4).

$$C_{HAADF} = \frac{I_{SiB}/V_{SiB}}{I_{Si\_}/V_{Si}} \qquad (4)$$

The integrated intensities are proportional to the atomic number Z at power 1.7 [12] and the volume of the Si:B lattice depends on the deformations measured in GPA.

$$I_{SiB} \propto x_B Z_B^{1.7} + (1-x_B)Z_{Si}^{1.7}$$
$$I_{Si\_} \propto Z_{Si}^{1.7}$$
$$V_{SiB} = a_{Si}^3 (1-\varepsilon_\perp)(1-\varepsilon\|)^2$$
$$V_{Si} = a_{Si}^3$$

Where the deformations of the two in plane directions are considered equal.
Thus, the B percent atomic concentration, $n_B$, can finally be calculated by:

$$n_{B\ HAADF}\ (\%) = \frac{[(1-\varepsilon_\perp)(1-\varepsilon\|)^2 C_{HAADF}-1]}{Z_B^{1.7}-Z_{Si}^{1.7}} \qquad (5)$$

Acknowledgments

We acknowledge fruitful discussions with L. Largeau, D. Bouchier, and A. van Waveren. This work was partly supported by the French CNRS RENATECH network. We acknowledge support from the French National Research Agency (ANR) under Contract No. ANR-16-CE24-0016-01, ANR-19-CE47-0010-03 and ANR-22-QUA2-0002-02, and as part of the "Investissements d'Avenir" program (Labex NanoSaclay, ANR-10-LABX-0035).